\documentclass[%
 aip,
 rsi,
 amsmath,amssymb,
 reprint,%
]{revtex4-1}

\usepackage{graphicx}% Include figure files
\usepackage{dcolumn}% Align table columns on decimal point
\usepackage{bm}% bold math
\usepackage[utf8]{inputenc}
\usepackage[T1]{fontenc}
\usepackage{mathptmx}
\usepackage{xcolor}
\usepackage{multirow}
\usepackage[english]{babel}
\usepackage{natbib}
\usepackage[breaklinks=true]{hyperref}
\usepackage{breakcites}

\begin{document}
	
\preprint{AIP/123-QED}
	
\title[A Method for Reducing the Adverse Effects of Stray-Capacitance on Capacitive Sensor Circuits]{A Method for Reducing the Adverse Effects of Stray-Capacitance on Capacitive Sensor Circuits}
% Force line breaks with \\
	
\author{C. Gettings}
\author{C. C. Speake}%
\affiliation{ 
Astrophysics and Space Research Group, School of Physics and Astronomy, University of Birmingham, Edgbaston, Birmingham, B15 2TT, United Kingdom%\\This line break forced with \textbackslash\textbackslash
}%
	
\date{\today}% It is always \today, today,

\begin{abstract}
We examine the increase in voltage noise in capacitive sensor
circuits due to the stray-capacitance introduced by connecting cables.
We have measured and modelled the voltage noise of various standard
circuits, and we compare their performance against a benchmark without
stray-capacitance that is optimised to have a high signal-to-noise
ratio (SNR) for our application. We show that a factor limiting sensitivity
is the so-called noise gain, which is not easily avoided. In our
application the capacitive sensor is located in a metallic vessel
and is therefore shielded to some extent from ambient noise at radio
frequencies. It is therefore possible to compromise the shielding
of the coaxial connecting cable by effectively electrically floating
it. With a cable stray-capacitance of $1.8nF$ and at a modulation
frequency of $100kHz$, our circuit has an output voltage noise a
factor of 3 larger than the benchmark. 
\end{abstract}

\maketitle

\section{Introduction}
\label{Introduction}

Capacitive sensors have a long history of use in scientific measurement,
\cite{Origin} and are used extensively across various branches of
science. \cite{Uses} Applications include displacement transducers
\cite{DispMeas} or straightforward capacitance meters. \cite{ModStray}
See the recent review article by Ramanathan et al. for further examples of capacitive sensors. \cite{New}
Capacitive displacement sensors are attractive devices as they have
no Johnson noise and their sensitivity is therefore limited by the
noise from the electronics that provides them with a measurable readout.
However, for reasonable geometries and frequencies, capacitive sensors
have large impedances, and this leads to them being sensitive to the
stray-capacitances of coaxial cables that are often needed to connect
them to a pre-amplifier. The literature often describes the problems
associated with stray-capacitances in terms of the biases they create
rather than with increase in noise. This paper discusses the latter
issue. As discussed below, the stray-capacitance can increase the
noise gain and severely reduce the signal-to-noise ratio (SNR). One
way to mitigate the effects of stray-capacitance is to reduce the
input impedance of the sensor by using a resonant circuit
or an impedance matching transformer.\cite{Accel} However, these techniques necessarily introduce Johnson noise. Further, stray-capacitance associated with
the required inductances, transformers, and circuit boards can shift
the resonant frequency and make these strategies less
robust and straightforward.\cite{Board} This paper proposes an alternative
viable method that limits the effect of stray-capacitance on the noise
gain as explained below. This improvement comes with a large reduction
in the effectiveness of the coaxial cable in shielding the device
from ambient electromagnetic environmental interference (RFI). However
in many sensitive applications the sensor is placed in a metallic
vessel, such as a vacuum chamber or low temperature cryogenic dewar (our case),
which can provide reasonable shielding from RFI and additional shielding
from coaxial cables may not be necessary. In such cases degradation
of the SNR is avoidable. The optimum choice comes as a trade-off between
required SNR and ambient noise.

Stray-capacitance can be a particular issue in cryogenic experimental
set-ups where it is necessary to use long coaxial cables that have
a small gauge in order to minimise their heat leak. However such fine
coaxial cables have large capacitance per unit length. In practice
lock-in amplifiers operating at specific frequencies can be used to
avoid the effects of interference out of the demodulation band, which
can typically be at around 100 $kHz$. This technique can be successfully
employed provided the dynamic range of the pre-amplifier in the detection
circuit is not exceeded by the RFI. Therefore, in addition to the
noise performance, the effectiveness of the coaxial cable and the
surrounding vessel in rejecting electromagnetic interference must
be taken into consideration. We suggest below that, by making a minor
adjustment to a typical transimpedance capacitive sensor circuit,
a satisfactory compromise can be obtained between having a usable
noise performance and an acceptable rejection of RFI. We present measurements
taken in our laboratory to support this.

\section{Noise Gain}

\begin{figure}
	\includegraphics[width=1\columnwidth]{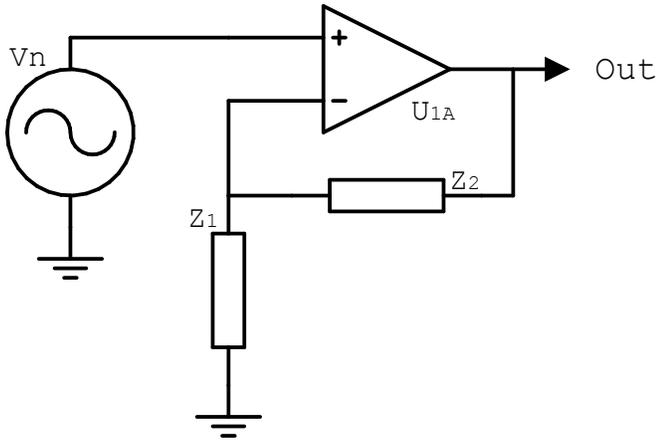} \caption{A diagram of a typical non-inverting amplifier where the voltage noise of the amplifier is represented by a voltage generator at the non-inverting input.}
	\label{Benchmark} 
\end{figure}

\begin{figure}
	\includegraphics[width=1\columnwidth]{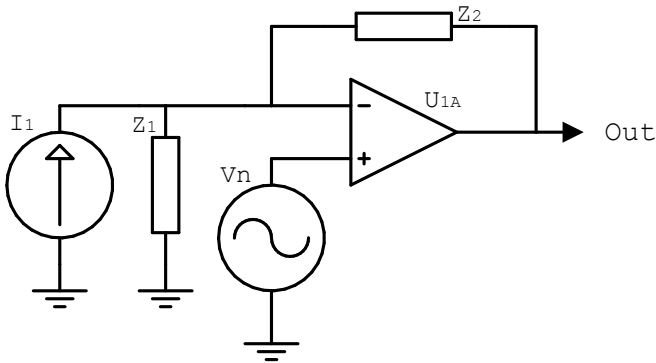}
	\caption{A diagram of a typical transimpedance circuit: taking a
		current input signal $I_{1}$ and converting this to a voltage output signal by the op-amp labelled $U_{1A}$.}
	\label{Benchmark2} 
\end{figure}

Before we proceed to describe the details of the circuits that we
have tested we will explicitly define the problem of the excess noise
introduced by stray-capacitance at the input of an amplifier. Noise gain is a term used in the design of operational amplifiers
and appears, from conversations with physicists and engineers, to
be little appreciated. However for our purposes it has a definite
simple meaning that is illustrated by Figures~\ref{Benchmark} and~\ref{Benchmark2}. In Figure~\ref{Benchmark} 
we show a simple diagram of a non-inverting amplifier were we have represented its voltage
noise, $V_{n}$, in the usual way as a random voltage generator at
the non-inverting input of the amplifier. It is clear that the output voltage noise is~\cite{AmpRef}
\begin{equation}
V_{o}=GV_{n},\label{eq:1a}
\end{equation}
where 
\begin{equation}
G=1+\frac{Z_{2}}{Z_{1}}.\label{eq:1b}
\end{equation}
$G$ here is the noise gain but it is easily
seen to also be the non-inverting signal gain of the amplifier
in the infinite gain bandwidth approximation. Now if we examine Figure~\ref{Benchmark2}
where we have redrawn the amplifier in a transimpedance mode with a high
impedance current source, $I_{1}$, with impedance, $Z_{1}$, connected to its inverting input. It is clear that the output voltage noise from the op-amp is again given by Equation \ref{eq:1a}. It is also clear that the presence of a low impedance to ground, such as that due to stray cable capacitance, at the inverting input of
the amplifier can introduce significant output voltage noise. 

\section{Circuit Tests}

\begin{figure}
	\includegraphics[width=1\columnwidth]{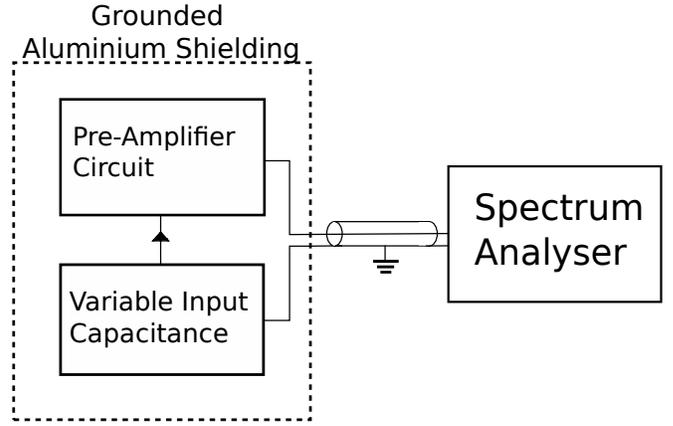}
	\caption{A schematic diagram of the measurement set-up for the circuit noise measurements.}
	\label{SchemMeas} 
\end{figure}

In this section we present the noise measurements of various capacitive sensor pre-amplifier circuits. The exact circuit used in each case is presented where relevant but the general schematic diagram for all measurements is shown in Figure~\ref{SchemMeas}, where the signal path to and from the spectrum anlyser used to measure the noise is shown.

In this section the variable input capacitance and pre-amplifier circuits are located on the same circuit board and within the same aluminium shielding. In Section~\ref{ShEff}, however, these two parts are separated as described in that section.

\subsection{Benchmark Circuit}

\begin{figure}
	\includegraphics[width=1\columnwidth]{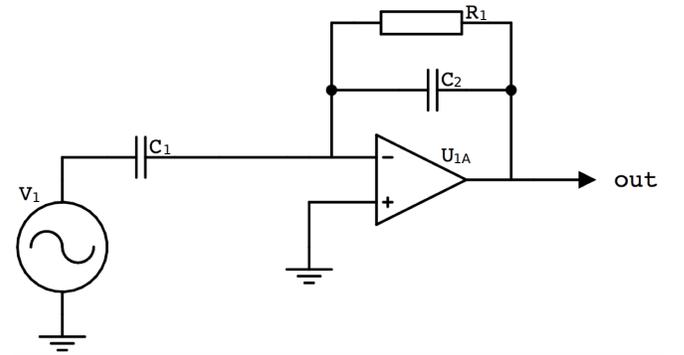}
	\caption{A diagram of the benchmark transimpedance circuit:
		taking a current input signal from the changing capacitance labelled
		$C_{1}$ that is driven by a voltage generator $V_{1}$. This is converted to a
		voltage output signal by the op-amp labelled $U_{1A}$.}
	\label{BenchmarkNoiseModel} 
\end{figure}

A standard transimpedance circuit acting as a capacitive sensor is
shown in Figure~\ref{BenchmarkNoiseModel}
where the voltage generator, $V_{1}$, connected to a changing capacitance $C_{1}$, produces a current input signal for the op-amp $U_{1A}$. The absence of stray-capacitance means that this circuit can be considered to be a benchmark in terms of noise. The feedback network consists of a resistor $R_{1}$ and capacitor $C_{2}$ in parallel.

The output voltage noise from this circuit can be modelled by taking
into consideration the main sources of noise present. These are the
thermal noise from the feedback resistor $R_{1}$, $V_{R1}$, the
current noise density of the op-amp being used, $I_{n}$, and the
op-amp's voltage noise density, $V_{n}$, which is multiplied by the
noise gain of the circuit, $G$.

Noise gain in general can be changed without
modifying signal gain. The noise gain for this benchmark circuit
is described by the following equation 
\begin{equation}
G=\frac{i\omega C_{1}R_{1}}{1+i\omega C_{2}R_{1}}+1,\label{SignalGain}
\end{equation}
where as usual $i$ is $\sqrt{(-1)}$. The total output voltage noise, $V_{T}$ ($\frac{V}{\sqrt{Hz}}$),
is then given by the equation~\cite{AmpRef}
\begin{equation}
V_{T}=\sqrt{(V_{R1}F)^{2}+(I_{n}R_{1})^{2}+(V_{n}G)^{2}},\label{VoltNoise}
\end{equation}
where the amplitude spectral density of Johnson noise in the feedback
resistor $R_{1}$ is given by the usual expression~\cite{AoE} 
\begin{equation}
V_{R1}=\sqrt{4k_{B}TR_{1}}.\label{eq:4}
\end{equation}
$F$ is a one-pole low-pass
filter term created by resistor $R_{1}$ and capacitor $C_{2}$ which
acts to attenuate the Johnson noise produced by resistor $R_{1}$
and any signal at frequencies above the pole of the filter $\left(1/2\pi R_{1}C_{2}\right)$.
In our measurements $C_{2}$ was chosen to be $1.8pF$, but in practice
the capacitance measured across $R_{1}$ varied between $3pF$ and
$4.5pF$ due to the capacitances present on the circuit board used, leading to the reduction in measured noises at higher frequencies.
The models used in this paper reflect these measured values. This
simple analysis ignores the effects of the finite bandwidth of the
op-amp, where the open-loop gain reduction at high frequencies must
be taken into account. This, however, is acceptable in this case as
the noise gain here is not affected by the smaller open-loop gain
at high frequencies. \cite{WJ}

\begin{figure}
\includegraphics[width=1\columnwidth]{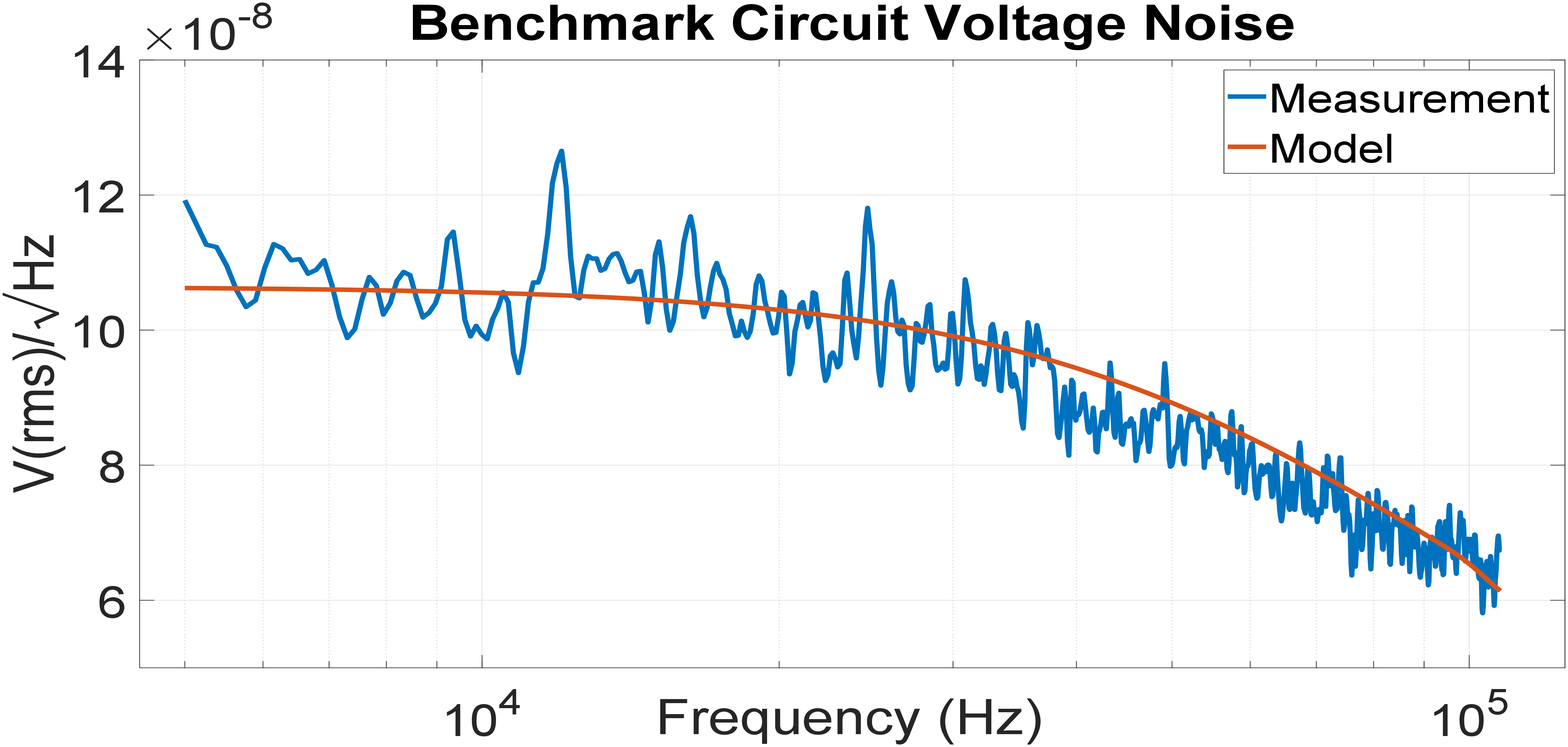} \caption{The voltage noise measurement of the benchmark circuit in Figure~\ref{BenchmarkNoiseModel}
and associated analytical noise model from Equations~\ref{SignalGain}
and~\ref{VoltNoise} using the same components as the measurement.}
\label{BenchmarkNoise} 
\end{figure}

The analytical expression for the voltage noise of this circuit in
Equation~\ref{VoltNoise} was compared to a direct measurement. In
the voltage noise measurement, and all other noise measurements described
in this paper, a Hewlett-Packard 35665A Dynamic Signal Analyser was
used to measure the voltage noise directly from the output of the
circuits. In order to eliminate as far as possible environmental background
noise, the circuits under test were contained in aluminium boxes of
wall thickness's of approximately $5mm$ and were powered by batteries.
The circuit was designed to optimise the noise performance for our
displacement transducer ignoring the
effects of any stray-capacitances.\cite{Float} This leads to a design with low
current noise and therefore large values of feedback resistance $R_{1}$.
The following component values were chosen for the benchmark circuit:
capacitor $C_{1}$ was $3.5pF$, $C_{2}$ was $1.8pF$, the resistor
$R_{1}$ was $620k\varOmega$ and the op-amp used for $U_{1A}$ was
a LTC6244HV. These values approximately match those that will be used
in the detection circuitry for the capacitive sensor in the experiment where the modulation frequency is set to $100kHz$.\cite{Float}
The measurement and model are compared in Figure~\ref{BenchmarkNoise}.
At $100kHz$ the voltage noise is approximately $67.9\frac{nV}{\sqrt{Hz}}$.

It is interesting to note that there is a peak in the noise plot in both Figure~\ref{BenchmarkNoise}, and later in Figure~\ref{GoodCircuitNoise}, at about 12kHz and must be due to some residual leakage of RFI into the set-up shown in Figure~\ref{SchemMeas}. Our models only look at the voltage noise inherent in the circuits alone as the Equations~\ref{SignalGain}-\ref{eq:4} suggest, and so we are not expecting to be able to predict and model the local interference in our laboratory. Despite this, the simple analytical model describes the noise well.

\subsection{Stray-capacitance Circuit}

\begin{figure}
\includegraphics[width=1\columnwidth]{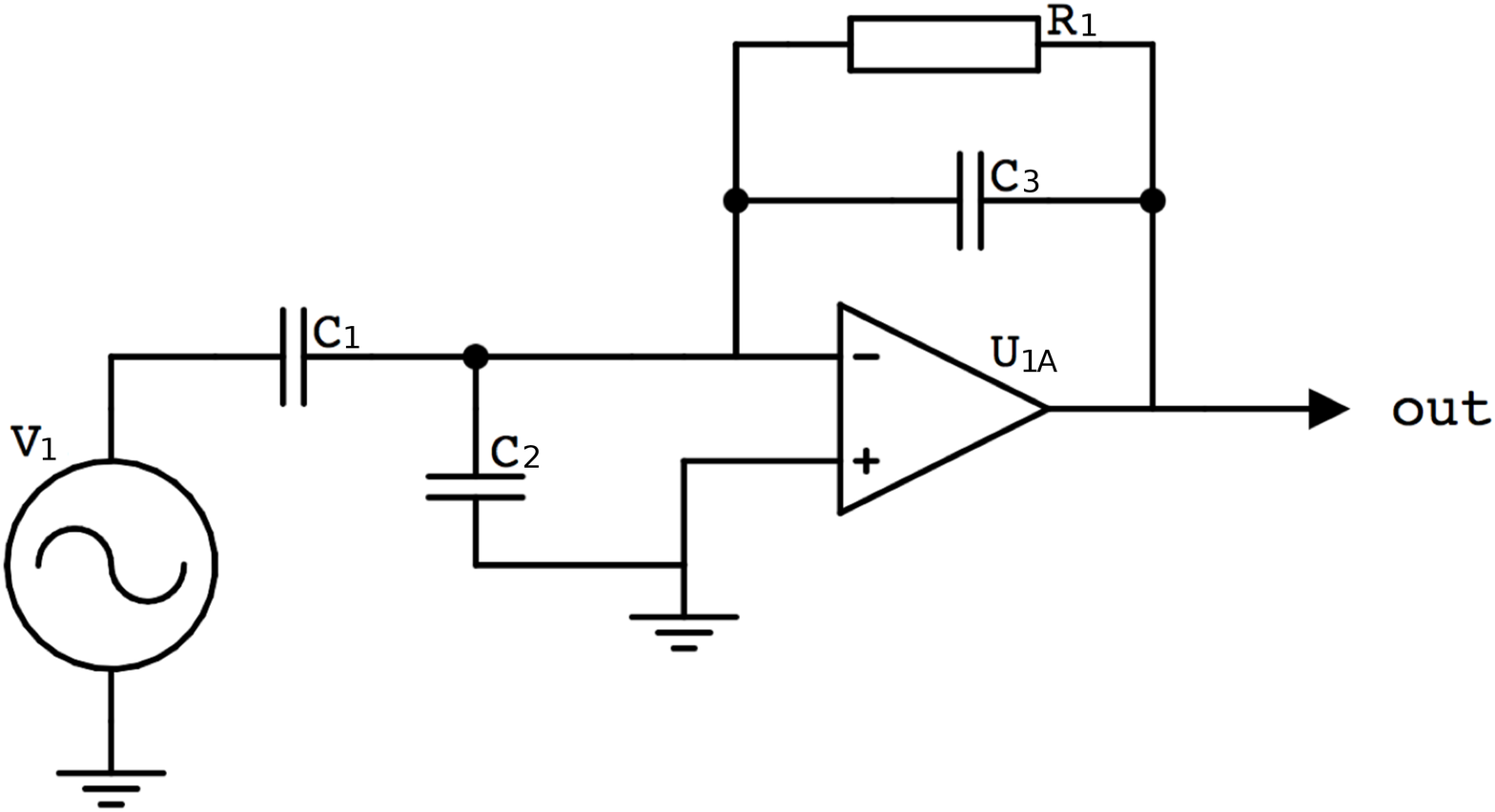} \caption{A diagram of a circuit (circuit B) which models a stray-capacitance
through the capacitor labelled $C_{2}$.}
\label{BadCircuit} 
\end{figure}

\begin{figure}
\includegraphics[width=1\columnwidth]{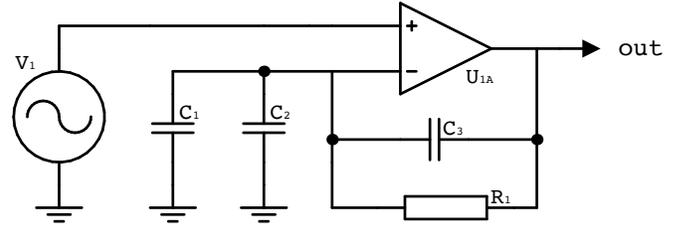} \caption{A diagram of the equivalent noise circuit from Figure~\ref{BadCircuit},
where the numbered components remain the same.}
\label{BadNoiseModel} 
\end{figure}

As a first attempt to model the effect of the stray-capacitance due
to a cable we use a capacitor, labelled $C_{2}$, which is connected
to ground as shown in Figure~\ref{BadCircuit}.
We refer to this circuit as circuit B. The noise gain for circuit
B is altered from that of the circuit in Figure~\ref{BenchmarkNoiseModel}
as the input impedance now comes from capacitors $C_{1}$ and $C_{2}$
in \textit{parallel}. An equivalent noise circuit for the noise gain
in Figure~\ref{BadCircuit} is shown in Figure~\ref{BadNoiseModel}.

\begin{figure}
\includegraphics[width=1\columnwidth]{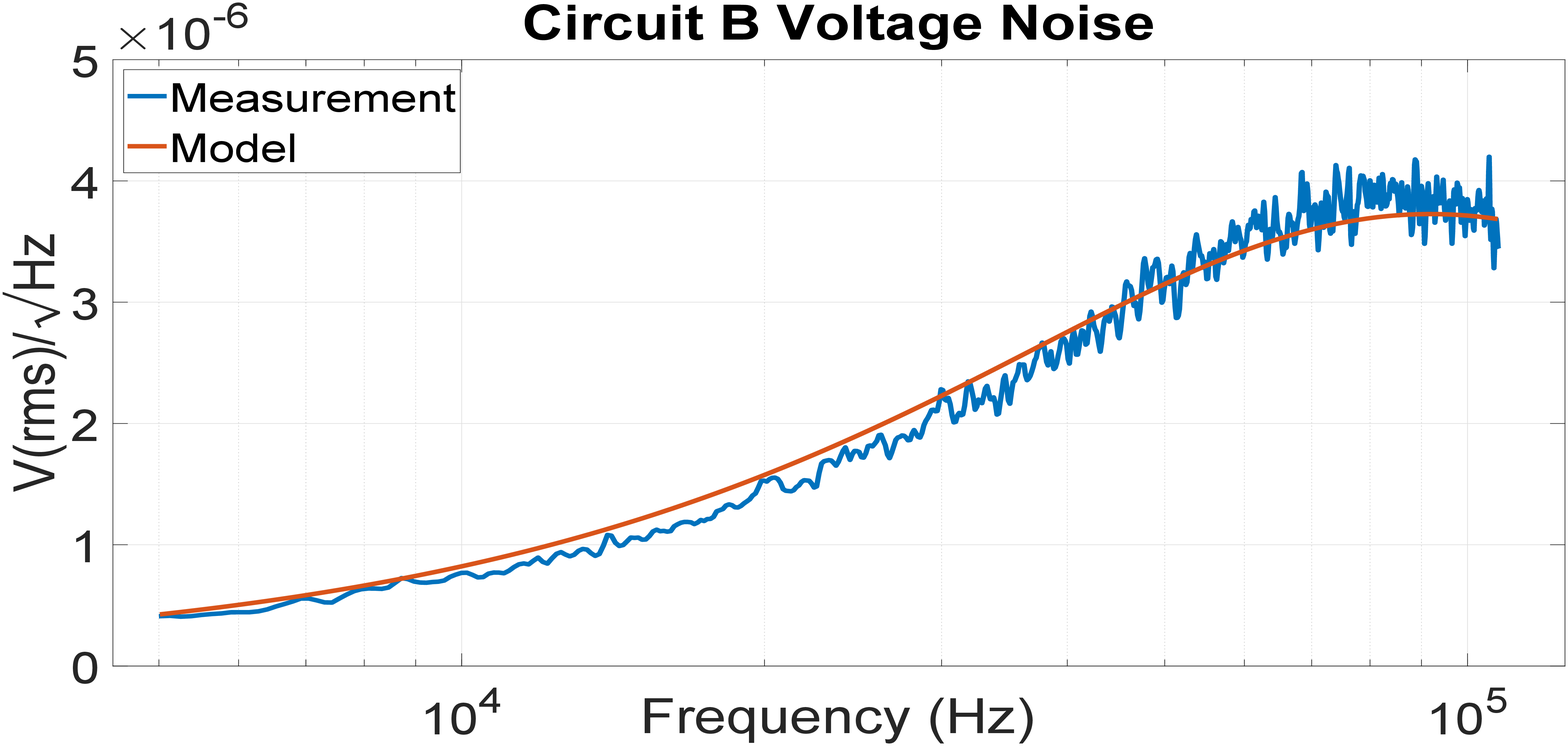} \caption{The voltage noise measurement of the stray-capacitance modelling circuit
(Circuit B) in Figure~\ref{BadCircuit} and associated analytical
noise model from Equations~\ref{VoltNoise} and~\ref{BadNG} using
the same components as the measurement.}
\label{BadCircuitNoise} 
\end{figure}

The noise gain for the circuit in Figure~\ref{BadCircuit} is given
by the following 
\begin{equation}
G=\frac{i\omega R_{1}(C_{1}+C_{2})}{1+i\omega C_{3}R_{1}}+1.\label{BadNG}
\end{equation}
In general $C_{2}>>C_{1}$ and this leads to a large increase in noise
gain. Inserting this new noise gain into Equation~\ref{VoltNoise} gives
an analytical expression for the voltage noise for circuit B. Unlike
the case of the benchmark circuit in Figure~\ref{BenchmarkNoiseModel}, the
finite bandwidth of the op-amp was taken into account here \cite{WJ}.
This predicted noise was compared to direct measurement using the
method mentioned previously. In measuring the noise of circuit B,
repeated components from Figure~\ref{BenchmarkNoiseModel} were kept the same
and $C_{2}$ was $1.8nF$ which matched approximately the stray-capacitance
of coaxial cable of the type used in our experiment at a length of
$3.5m$. The measurement and model are compared in Figure~\ref{BadCircuitNoise}.
At $100kHz$ the voltage noise is approximately $3.8\frac{\mu V}{\sqrt{Hz}}$.
The simple analytical model describes the noise well.

\subsection{Buffer Op-Amp Feedback Circuit}

\begin{figure}
\includegraphics[width=1\columnwidth]{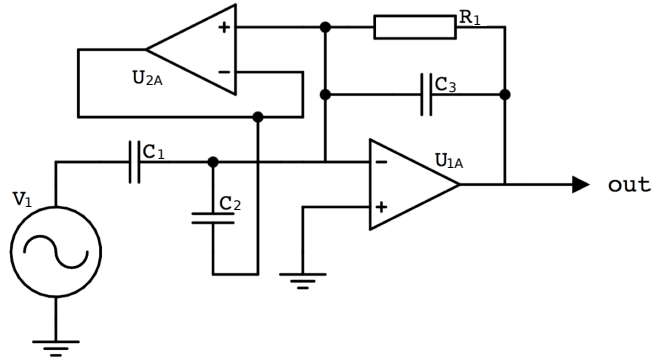} \caption{A diagram of a circuit (circuit C) using the common feedback technique
of a buffer op-amp, for reducing the effects of stray-capacitance
from the capacitor $C_{2}$.}
\label{Standard} 
\end{figure}

A common method for reducing the effects of stray-capacitance
is shown by the circuit in Figure~\ref{Standard}
labelled circuit C.\cite{Standard1,Standard2} In this situation a buffer op-amp, labelled $U_{2A}$, is used in a feedback loop to drive both sides of the capacitor, $C_{2}$, to the same voltage. In principle this should remove the effects of
stray-capacitance. While this method has been shown to eliminate the
bias capacitance due to the cable and the instability in circuits
with large stray-capacitances, depending on the choice of op-amps used,
\cite{Stable1,Stable2} it does not improve noise performance.

The output noise of circuit C was measured using the method mentioned
previously; the component values and types were identical to those
in Figure~\ref{BadCircuit}. The buffer op-amp $U_{2A}$ was initially
chosen to be an OP07. An analytical model for the voltage noise at
the output of this circuit can be derived by considering the noise
introduced across capacitor $C_{2}$ from the op-amp $U_{2A}$ in
Figure~\ref{Standard}. The total voltage noise, $V_{2T}$ ($\frac{V}{\sqrt{Hz}}$),
introduced is given by the equation 
\begin{equation}
V_{2T}=\frac{V_{2n}R_{1}F}{Z_{s}}\label{CurrentNoise2}
\end{equation}
where $V_{2n}$ is the buffer op-amp's voltage noise in units of ($\frac{V}{\sqrt{Hz}}$),
and $Z_{s}$ is the impedance of capacitor $C_{2}$. $R_{1}$ is the
resistance value in the circuit and $F$ is the low-pass filter term
from Equation~\ref{VoltNoise} which at $100kHz$ is 0.55. In practice the pole of $F$ would be at higher frequencies where it would not attenuate any signal.

\begin{figure}
\includegraphics[width=1\columnwidth]{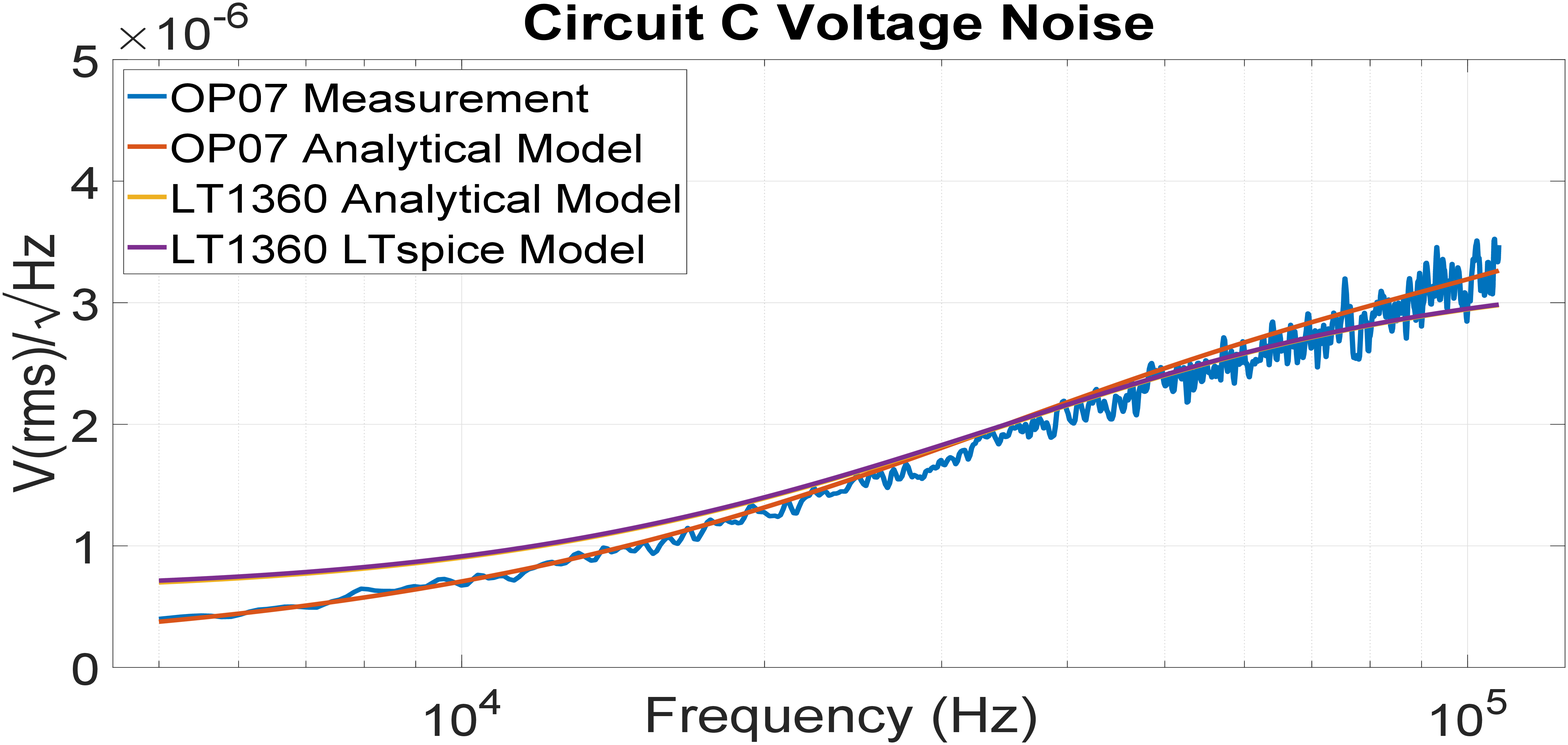} \caption{The voltage noise measurement of Circuit C in Figure~\ref{Standard}
and the associated noise model from Equation~\ref{CurrentNoise2}.
Also included are analytical and LTspice models using a LT1360 op-amp
rather than the OP07.}
\label{StandardNoise} 
\end{figure}

The voltage noise measurement and model are shown in Figure~\ref{StandardNoise}.
At $100kHz$ the voltage noise is approximately $3.0\frac{\mu V}{\sqrt{Hz}}$.
Also shown in Figure~\ref{StandardNoise} is the analytical model
and LTspice result for the LT1360, as $U_{2A}$, which is an operational
amplifier with a faster slew rate and lower voltage noise density
than the OP07. Using this op-amp the voltage noise is reduced to approximately
$2.9\frac{\mu V}{\sqrt{Hz}}$ at $100kHz$. The analytical model describes both the measured and LTspice predicted noise spectra well.

This analysis shows that the presence of the buffer op-amp $U_{2A}$
can contribute a significant amount of output voltage noise to the circuit,
as the stray-capacitance, $C_{2}$, will couple the noise to the signal
carrying core. \cite{ShieldNoise} It also suggests that the similarity
between the noise measurements in Figures~\ref{BadCircuitNoise}
and~\ref{StandardNoise} is coincidental. Where for Figure~\ref{BadCircuitNoise}
the increased noise comes from the op-amp voltage noise being multiplied
by the large noise gain, and for Figure~\ref{StandardNoise} it arises
from the feedback op-amp voltage noise being introduced across the
stray-capacitance.

\subsection{Proposed Noise Gain Modification Method}

Figure~\ref{StandardNoise} shows clearly that the conventional solution
to the problem of stray-capacitance, which although allows a circuit
to operate in a stable and accurate manner, gives an increased voltage
noise. The goal is then to reduce the effects of stray-capacitance
on the stable operation of a transimpedance circuit, while maintaining
as low a circuit noise as possible compared to the benchmark circuit.

\begin{figure}
\includegraphics[width=1\columnwidth]{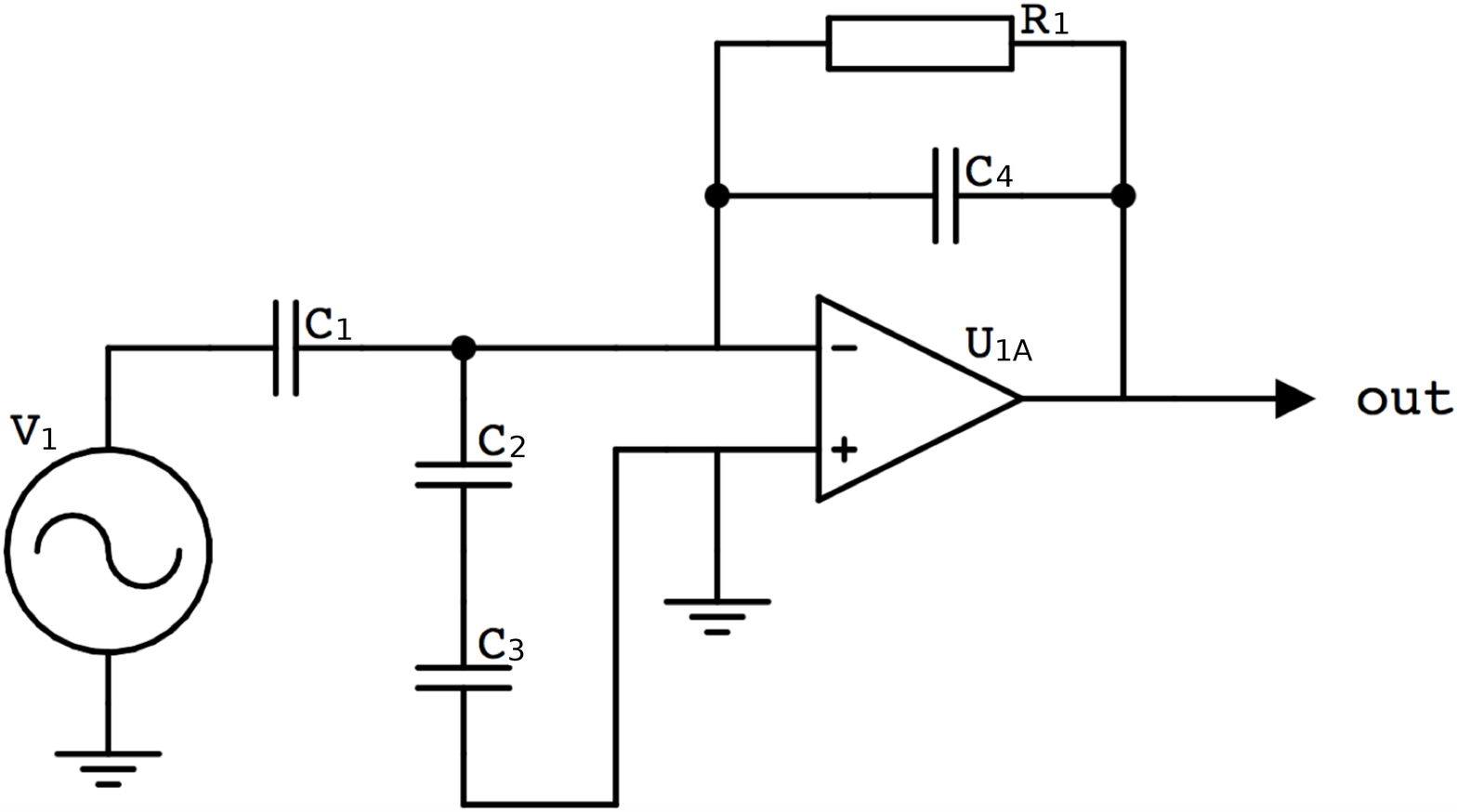} \caption{A diagram of a test circuit (circuit D) which modifies the noise gain
to offset the effects of the stray-capacitance from capacitor $C_{2}$.}
\label{GoodCircuit} 
\end{figure}

\begin{figure}
\includegraphics[width=1\columnwidth]{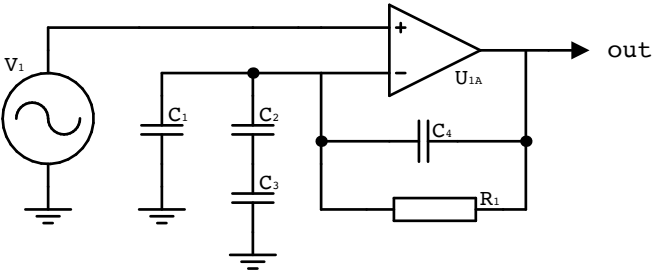} \caption{A diagram of the equivalent noise circuit from Figure~\ref{GoodCircuit},
where the numbered components remain the same.}
\label{GoodNoiseModel} 
\end{figure}

One way to do this is to reduce the noise gain of the circuit. This
can be done by using the set-up shown in Figure~\ref{GoodCircuit},
whose equivalent noise circuit for the noise gain is shown in Figure~\ref{GoodNoiseModel}.
In this circuit, labelled circuit D, the signal gain is the same as
for the benchmark circuit in Figure~\ref{BenchmarkNoiseModel}, but now the
noise gain is altered from Equation~\ref{BadNG} to the following
\begin{equation}
G=\frac{i\omega R_{1}(C_{1}(C_{2}+C_{3})+C_{2}C_{3})}{(1+i\omega C_{4}R_{1})(C_{2}+C_{3})}+1.\label{GoodNoiseGain}
\end{equation}
This change is due to the introduction of the capacitor labelled $C_{3}$,
which is in \textit{series} with the stray-capacitance $C_{2}$. If
$C_{3}<<C_{2}$, then Equation~\ref{GoodNoiseGain} reduces to Equation~\ref{SignalGain}
which is the benchmark best-case scenario.

\begin{figure}
\includegraphics[width=1\columnwidth]{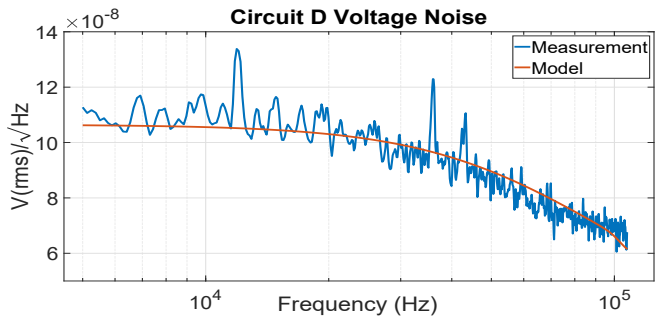} \caption{The voltage noise of the circuit where the noise gain has been modified
as shown in Figure~\ref{GoodCircuit} . Measured and predicted noise
from Equations~\ref{VoltNoise} and~\ref{GoodNoiseGain} using the
same components as the measurement. Notice that this is nominally
the same as Figure~\ref{BenchmarkNoise}.}
\label{GoodCircuitNoise} 
\end{figure}

When this new noise gain is used with the general noise formula given
in Equation~\ref{VoltNoise}, we obtain an expression for the voltage
noise of this new circuit. The noise of circuit D was measured using
the method mentioned previously. Repeated components from Figure~\ref{BadCircuit}
were as before and $C_{3}$ was $1.8pF$. The measurement and model
are compared in Figure~\ref{GoodCircuitNoise}.
At $100kHz$ the voltage noise is approximately $68.8\frac{nV}{\sqrt{Hz}}$.
As with the previous noise measurements, the model matches the measured
voltage noise well.

\section{Shield Effectiveness}
\label{ShEff}

For our candidate circuits it is important
to quantify the effectiveness of the coaxial cable in shielding the
signal carrying wire from environmental electromagnetic interference. This was done using circuits B and D, as circuit B is the typical shielded scenario and circuit D represents our method where the shield is now floating with respect to ground and as such is less effective. The in-situ cables in our experiment were attached
to the inputs of these two circuits whilst they were located in our
laboratory. This allowed a direct comparison of the output noise to
be made under lab conditions. A measurement was also made with the
benchmark circuit but with an unshielded cable at the input to the
op-amp, to give a reference noise for the background interference. A very important feature to note here is that the majority of the length of cables involved were inside a large metallic cryogenic dewar (which was electrically floating with respect to the circuits under test through a $1.8pF$ capacitor to ground) which would have provided extra shielding regardless of whether the coaxial shield was floated or not. This is typical for any cryogenic experiment and so is a fair comparison. 

\begin{figure}
\includegraphics[width=1\columnwidth]{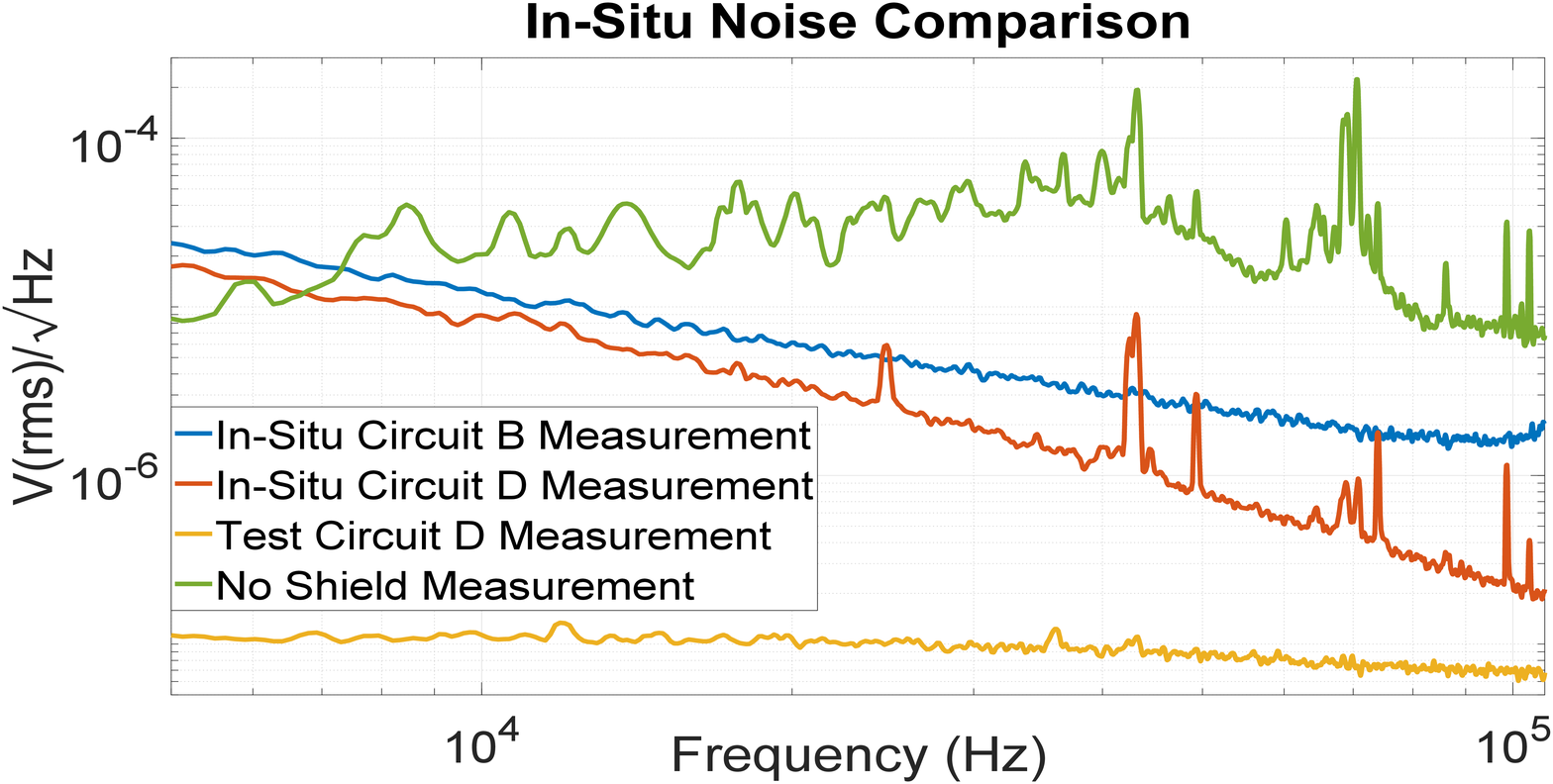} \caption{Real coaxial cable voltage noise measurements of circuits B and D
compared to an unshielded cable benchmark circuit and test circuit
D noise measurement.}
\label{RealCableNoise} 
\end{figure}

These measurements, along with the test measurement of circuit D from Figure~\ref{GoodCircuitNoise} for comparison, are shown in Figure~\ref{RealCableNoise}.
This figure shows that although there are significant interference noise peaks at certain frequencies for the in-situ circuit D, overall there is a much lower noise floor compared to the in-situ circuit B implying that the lowering of noise gain is worth the reduction in electromagnetic interference shielding. Also the no-shield scenario is shown to be a non-viable alternative to the problem of parasitic capacitance from shields. At $100kHz$ the in-situ circuit D noise is approximately 3.1 times larger than the test circuit D case. Aside from interference, additional noise creating this disparity between scenarios could arise from unaccounted for parasitic capacitances within the experimental set-up. As already stated, our set-up of course benefited from the fact that for the in-situ measurements, the majority of the cabling was contained inside a cryogenic dewar, which would have provided additional shielding.

\section{Summary}

\begin{table}[!htbp]
	\centering %
	\caption{A summary of the voltage noise results at $100kHz$ for the
		benchmark circuit, and circuits B, C, D. The in-situ measurements are also included.}
	\begin{ruledtabular}
		\begin{tabular}{l c r}
			Circuit & Voltage Noise $(\frac{nV}{\sqrt{Hz}})$ &  Disp. Noise  $(\frac{pm}{\sqrt{Hz}})$\\
			\hline 
			Benchmark Circuit  & $67.9$ & $6.2$ \\
			Circuit B  & $3767.0$ & $341.5$ \\
			Circuit C OP07 & $3048.0$ & $276.3$ \\
			Circuit C LT1360  & $2940.0$ & $266.5$ \\
			Circuit D  & $68.8$ & $6.2$ \\
			In-Situ Circuit B & $1665.0$ & $150.9$ \\
			In-Situ Circuit D & $214.8$ & $19.5$ \\
			In-Situ No-Shield & $8059.0$ & $730.5$ \\
		\end{tabular}
		\label{Summary}
	\end{ruledtabular}
\end{table}
A summary of the voltage noises at $100kHz$ for the benchmark circuit,
and circuits B, C and D from Figures~\ref{BenchmarkNoiseModel},~\ref{BadCircuit},~\ref{Standard}
and~\ref{GoodCircuit} respectively, is listed in Table~\ref{Summary}.
The same is done for the in-situ circuits B and
D as well as the no-shield circuit from Figure~\ref{RealCableNoise} for comparison.

These noise values have been converted into displacement noises in the table. This is to highlight the projected physical sensitivities of these circuits based on the specifications of our experiment, where we have a bias AC drive voltage of $4.0V_{pp}$, a capacitor plate separation of the order of $1mm$ in the capacitive transducer bridge and a capacitive plate area of $2 \times 10^{-4}m^{2}$.

\section{Discussion}

The voltage noise from a variety of pre-amplifier circuits
for non-resonant capacitive sensors has been measured and modelled
in order to explore the negative effects of large stray-capacitances
at their inputs. Five different test circuits
were compared: one was a standard pre-amplifier that was optimised
for our application with no stray-capacitance (Benchmark circuit);
one was the standard pre-amplifier but now with stray-capacitance
(circuit B); another two employed active feedback to maintain the
cable shield at the voltage of the input to the op-amp (circuit C
using two different amplifiers for the active feedback). Finally, a novel circuit (circuit D) was tested. A summary of the noise
characteristics of these circuits is shown in Table~\ref{Summary}.
Circuit D had the best noise characteristics, along with the benchmark circuit, and is shown in Figure~\ref{GoodCircuit}. The
addition of a simple small capacitance, $C_{3}$, connecting the
shield to the non-inverting input of the pre-amplifier reduces the
noise gain that is produced by the low impedance of the shield. The
voltage noise for this circuit is shown in Figure~\ref{GoodCircuitNoise}. 

In-situ results indicate a modest increase in noise, by a factor of 3, due to environmental interference and the non-static capacitances present in our actual set-up.
This increase is, however, still acceptable at the signal modulation frequency of $100kHz$. This indicates environmental interference is relatively
benign in this situation and so the reduced shield effectiveness is
not significant in this case. In our experiment this is due to the
fact that the signal carrying coaxial cable is mostly inside a cryogenic
dewar which is effectively a Faraday cage and a shield against high
frequency magnetic fields. Clearly standard modulation/demodulation
schemes can mitigate against low frequency interference if it does
not overwhelm the pre-amplifier. The degradation
in the measured shielding effectiveness is not equivalent to there being
no shield at all, as shown in Figure~\ref{RealCableNoise}.

\section{Conclusion}

The proposed circuit is a trivial extension of the
usual methods of designing a capacitive sensor circuit. However, as far as we are aware, it has not been proposed
previously most likely due to the perception that it is incapable
of providing shielding from environmental interference. The proposed
technique could be successfully employed in many applications in addition to cryogenic experiments. For
example, vacuum systems are employed in many sensitive experiments
and would provide electromagnetic shielding from external noise sources.
In general, however, there may be situations were interference is
a dominant factor that determines a capacitive sensor's sensitivity.
A particularly important case is when an unshielded drive signal
at the input of a bridge circuit ($V_{1}$ in Figures~\ref{BenchmarkNoiseModel},~\ref{BadCircuit},~\ref{Standard} and~\ref{GoodCircuit})
can couple directly to the pre-amplifier. In such cases a method for
which the shield effectiveness is preserved should be employed. 

One way to balance the benefits of a decrease in noise gain with the negatives of a reduction in shield effectiveness may be to make capacitor $C_{3}$ from Figure~\ref{GoodCircuit} a variable capacitor. This could allow the trade-off to be fine-tuned for more general applications, and may be a more general
solution to the problem of stray-capacitance when external interference
cannot be ignored and where fine cables are required to minimise
the heat leak in a cryogenic experiment. In our case, placing the
pre-amplifier circuit near the point where the coaxial cable
leaves the cryogenic dewar further reduced interference and drove
the voltage noise closer to the ideal value.

Finally, it may be useful to the reader if we compare the performance of the capacitive sensor presented here with, for example, a resonant capacitive sensor. \cite{Accel} In general, the sensitivity of capacitive sensors is proportional to the bias AC drive voltage applied to the electrodes, and inversely proportional to the equilibrium separation distance between the two sensing electrodes. For the resonant bridge transducer these parameters are $2.1V_{pp}$ and $0.25mm$ respectively.\cite{Accel} Other factors such as capacitance plate area and drive frequency are important but strongly depend on each individual case as to what is practical. From Table I, the in-situ circuit D sensor has a displacement noise of $19.5\frac{pm}{\sqrt{Hz}}$ that would be expected to be about $14\frac{pm}{\sqrt{Hz}}$ after coherent demodulation, whereas the resonant bridge transducer has a noise floor of approximately $1.0\frac{pm}{\sqrt{Hz}}$ above $1Hz$. \cite{Accel} We can estimate an approximate sensitivity of our detector if we employed the same gap and drive voltage as that for the resonant sensor. These parameters can be easily modified in our design. We then find that our device would be a factor of about 7 less sensitive. Nevertheless we believe the simplicity of the transimpedance scheme that we have presented here has its merits. Further our device has been demonstrated in a cryogenic setting and has sufficient sensitivity for our application. \cite{Float}

\begin{acknowledgments}
We would like to thank David Hoyland for many useful and constructive
discussions. We are grateful to UK STFC (grant number: ST/F00673X/1)
who initially supported this work. We are also very grateful to Leverhulme
(grant number: RPG-2012-674) for financial support. We would also
like to Stefan Schlamminger for refocussing our efforts toward a capacitive
sensing solution for our experiment. 
\end{acknowledgments}

\nocite{*}
\bibliography{aipsamp}% Produces the bibliography via BibTeX.

\providecommand{\noopsort}[1]{}\providecommand{\singleletter}[1]{#1}%
\begin{thebibliography}{16}%
\makeatletter
\providecommand \@ifxundefined [1]{%
 \@ifx{#1\undefined}
}%
\providecommand \@ifnum [1]{%
 \ifnum #1\expandafter \@firstoftwo
 \else \expandafter \@secondoftwo
 \fi
}%
\providecommand \@ifx [1]{%
 \ifx #1\expandafter \@firstoftwo
 \else \expandafter \@secondoftwo
 \fi
}%
\providecommand \natexlab [1]{#1}%
\providecommand \enquote  [1]{``#1''}%
\providecommand \bibnamefont  [1]{#1}%
\providecommand \bibfnamefont [1]{#1}%
\providecommand \citenamefont [1]{#1}%
\providecommand \href@noop [0]{\@secondoftwo}%
\providecommand \href [0]{\begingroup \@sanitize@url \@href}%
\providecommand \@href[1]{\@@startlink{#1}\@@href}%
\providecommand \@@href[1]{\endgroup#1\@@endlink}%
\providecommand \@sanitize@url [0]{\catcode `\\12\catcode `\$12\catcode
  `\&12\catcode `\#12\catcode `\^12\catcode `\_12\catcode `\%12\relax}%
\providecommand \@@startlink[1]{}%
\providecommand \@@endlink[0]{}%
\providecommand \url  [0]{\begingroup\@sanitize@url \@url }%
\providecommand \@url [1]{\endgroup\@href {#1}{\urlprefix }}%
\providecommand \urlprefix  [0]{URL }%
\providecommand \Eprint [0]{\href }%
\providecommand \doibase [0]{http://dx.doi.org/}%
\providecommand \selectlanguage [0]{\@gobble}%
\providecommand \bibinfo  [0]{\@secondoftwo}%
\providecommand \bibfield  [0]{\@secondoftwo}%
\providecommand \translation [1]{[#1]}%
\providecommand \BibitemOpen [0]{}%
\providecommand \bibitemStop [0]{}%
\providecommand \bibitemNoStop [0]{.\EOS\space}%
\providecommand \EOS [0]{\spacefactor3000\relax}%
\providecommand \BibitemShut  [1]{\csname bibitem#1\endcsname}%
\let\auto@bib@innerbib\@empty
%</preamble>
\bibitem [{\citenamefont {Jones}\ and\ \citenamefont
  {Richards}(1973)}]{Origin}%
  \BibitemOpen
  \bibfield  {author} {\bibinfo {author} {\bibfnamefont {R.~V.}\ \bibnamefont
  {Jones}}\ and\ \bibinfo {author} {\bibfnamefont {J.~C.}\ \bibnamefont
  {Richards}},\ }\href {\doibase 10.1088/0022-3735/6/7/001} {\bibfield
  {journal} {\bibinfo  {journal} {Journal of Physics E-Scientific Instruments}\
  }\textbf {\bibinfo {volume} {6}},\ \bibinfo {pages} {589} (\bibinfo {year}
  {1973})}\BibitemShut {NoStop}%
\bibitem [{\citenamefont {Smith}\ and\ \citenamefont {Seugling}(2006)}]{Uses}%
  \BibitemOpen
  \bibfield  {author} {\bibinfo {author} {\bibfnamefont {S.~T.}\ \bibnamefont
  {Smith}}\ and\ \bibinfo {author} {\bibfnamefont {R.~M.}\ \bibnamefont
  {Seugling}},\ }\href {\doibase 10.1016/j.precisioneng.2005.10.003} {\bibfield
   {journal} {\bibinfo  {journal} {Precision Engineering-Journal of the
  International Societies for Precision Engineering and Nanotechnology}\
  }\textbf {\bibinfo {volume} {30}},\ \bibinfo {pages} {245} (\bibinfo {year}
  {2006})}\BibitemShut {NoStop}%
\bibitem [{\citenamefont {Chatterjee}\ \emph {et~al.}(2017)\citenamefont
  {Chatterjee}, \citenamefont {Mahato}, \citenamefont {Chattopadhyay},\ and\
  \citenamefont {De}}]{DispMeas}%
  \BibitemOpen
  \bibfield  {author} {\bibinfo {author} {\bibfnamefont {K.}~\bibnamefont
  {Chatterjee}}, \bibinfo {author} {\bibfnamefont {S.~N.}\ \bibnamefont
  {Mahato}}, \bibinfo {author} {\bibfnamefont {S.}~\bibnamefont
  {Chattopadhyay}}, \ and\ \bibinfo {author} {\bibfnamefont {D.}~\bibnamefont
  {De}},\ }\href {\doibase 10.1134/S0020441217010055} {\bibfield  {journal}
  {\bibinfo  {journal} {Instruments and Experimental Techniques}\ }\textbf
  {\bibinfo {volume} {60}},\ \bibinfo {pages} {154} (\bibinfo {year}
  {2017})}\BibitemShut {NoStop}%
\bibitem [{\citenamefont {Oven}(2014)}]{ModStray}%
  \BibitemOpen
  \bibfield  {author} {\bibinfo {author} {\bibfnamefont {R.}~\bibnamefont
  {Oven}},\ }\href {\doibase 10.1109/TIM.2014.2298673} {\bibfield  {journal}
  {\bibinfo  {journal} {IEEE Transactions on Instrumentation and Measurement}\
  }\textbf {\bibinfo {volume} {63}},\ \bibinfo {pages} {1748} (\bibinfo {year}
  {2014})}\BibitemShut {NoStop}%
\bibitem [{\citenamefont {Ramanathan}\ \emph {et~al.}(2016)\citenamefont
  {Ramanathan}, \citenamefont {Ramasamy}, \citenamefont {Jain}, \citenamefont
  {Nagrecha}, \citenamefont {Paul}, \citenamefont {Arulmozhivarman},\ and\
  \citenamefont {Tatavarti}}]{New}%
  \BibitemOpen
  \bibfield  {author} {\bibinfo {author} {\bibfnamefont {P.}~\bibnamefont
  {Ramanathan}}, \bibinfo {author} {\bibfnamefont {S.}~\bibnamefont
  {Ramasamy}}, \bibinfo {author} {\bibfnamefont {P.}~\bibnamefont {Jain}},
  \bibinfo {author} {\bibfnamefont {H.}~\bibnamefont {Nagrecha}}, \bibinfo
  {author} {\bibfnamefont {S.}~\bibnamefont {Paul}}, \bibinfo {author}
  {\bibfnamefont {P.}~\bibnamefont {Arulmozhivarman}}, \ and\ \bibinfo {author}
  {\bibfnamefont {R.}~\bibnamefont {Tatavarti}},\ }in\ \href@noop {}
  {{\selectlanguage {English}\emph {\bibinfo {booktitle} {Sensors, Transducers,
  Signal Conditioning and Wireless Sensors Networks}}}},\ \bibinfo {series}
  {Advances in Sensors-Reviews}, Vol.~\bibinfo {volume} {3},\ \bibinfo {editor}
  {edited by\ \bibinfo {editor} {\bibfnamefont {S.}~\bibnamefont {Yurish}}}\
  (\bibinfo  {publisher} {Int Frequency Sensor Assoc-IFSA},\ \bibinfo {address}
  {C/Esteve Terradas, Parc UPC-PMT, Edifici RDIT-KAM, 1, Barcelona,
  Castelldefels 08860, Spain},\ \bibinfo {year} {2016})\ pp.\ \bibinfo {pages}
  {213--227}\BibitemShut {NoStop}%
\bibitem [{\citenamefont {Bertolini}\ \emph {et~al.}(2006)\citenamefont
  {Bertolini}, \citenamefont {DeSalvo}, \citenamefont {Fidecaro}, \citenamefont
  {Francesconi}, \citenamefont {Marka}, \citenamefont {Sannibale},
  \citenamefont {Simonetti}, \citenamefont {Takamori},\ and\ \citenamefont
  {Tariq}}]{Accel}%
  \BibitemOpen
  \bibfield  {author} {\bibinfo {author} {\bibfnamefont {A.}~\bibnamefont
  {Bertolini}}, \bibinfo {author} {\bibfnamefont {R.}~\bibnamefont {DeSalvo}},
  \bibinfo {author} {\bibfnamefont {F.}~\bibnamefont {Fidecaro}}, \bibinfo
  {author} {\bibfnamefont {M.}~\bibnamefont {Francesconi}}, \bibinfo {author}
  {\bibfnamefont {S.}~\bibnamefont {Marka}}, \bibinfo {author} {\bibfnamefont
  {V.}~\bibnamefont {Sannibale}}, \bibinfo {author} {\bibfnamefont
  {D.}~\bibnamefont {Simonetti}}, \bibinfo {author} {\bibfnamefont
  {A.}~\bibnamefont {Takamori}}, \ and\ \bibinfo {author} {\bibfnamefont
  {H.}~\bibnamefont {Tariq}},\ }\href {\doibase 10.1016/j.nima.2006.04.041}
  {\bibfield  {journal} {\bibinfo  {journal} {Nuclear Instruments \& Methods in
  Physics Research Section A - Accelerators Spectrometers Detectors and
  Associated Equipment}\ }\textbf {\bibinfo {volume} {564}},\ \bibinfo {pages}
  {579} (\bibinfo {year} {2006})}\BibitemShut {NoStop}%
\bibitem [{\citenamefont {Hu}\ \emph {et~al.}(2014)\citenamefont {Hu},
  \citenamefont {Bai}, \citenamefont {Zhou}, \citenamefont {Li},\ and\
  \citenamefont {Luo}}]{Board}%
  \BibitemOpen
  \bibfield  {author} {\bibinfo {author} {\bibfnamefont {M.}~\bibnamefont
  {Hu}}, \bibinfo {author} {\bibfnamefont {Y.~Z.}\ \bibnamefont {Bai}},
  \bibinfo {author} {\bibfnamefont {Z.~B.}\ \bibnamefont {Zhou}}, \bibinfo
  {author} {\bibfnamefont {Z.~X.}\ \bibnamefont {Li}}, \ and\ \bibinfo {author}
  {\bibfnamefont {J.}~\bibnamefont {Luo}},\ }\href {\doibase 10.1063/1.4873334}
  {\bibfield  {journal} {\bibinfo  {journal} {Review of Scientific
  Instruments}\ }\textbf {\bibinfo {volume} {85}} (\bibinfo {year} {2014}),\
  10.1063/1.4873334}\BibitemShut {NoStop}%
\bibitem [{Amp(1992)}]{AmpRef}%
  \BibitemOpen
  \href@noop {} {\emph {\bibinfo {title} {Amplifier Reference Manual}}}\
  (\bibinfo  {publisher} {Analog Devices, Inc.},\ \bibinfo {year}
  {1992})\BibitemShut {NoStop}%
\bibitem [{\citenamefont {Horowitz}\ and\ \citenamefont {Hill}(1989)}]{AoE}%
  \BibitemOpen
  \bibfield  {author} {\bibinfo {author} {\bibfnamefont {P.}~\bibnamefont
  {Horowitz}}\ and\ \bibinfo {author} {\bibfnamefont {W.}~\bibnamefont
  {Hill}},\ }\href@noop {} {\emph {\bibinfo {title} {The Art of
  Electronics}}},\ \bibinfo {edition} {2nd}\ ed.\ (\bibinfo  {publisher}
  {Cambridge University Press},\ \bibinfo {year} {1989})\BibitemShut {NoStop}%
\bibitem [{\citenamefont {Jung}(1980)}]{WJ}%
  \BibitemOpen
  \bibfield  {author} {\bibinfo {author} {\bibfnamefont {W.~G.}\ \bibnamefont
  {Jung}},\ }\href@noop {} {\emph {\bibinfo {title} {IC Op-Amp Cookbook}}},\
  \bibinfo {edition} {2nd}\ ed.\ (\bibinfo  {publisher} {Howard W. Sams \& Co.,
  Inc.},\ \bibinfo {year} {1980})\BibitemShut {NoStop}%
\bibitem [{\citenamefont {Chalkley}\ \emph {et~al.}(2011)\citenamefont
  {Chalkley}, \citenamefont {Aston}, \citenamefont {Collins}, \citenamefont
  {Nelson},\ and\ \citenamefont {Speake}}]{Float}%
  \BibitemOpen
  \bibfield  {author} {\bibinfo {author} {\bibfnamefont {E.~C.}\ \bibnamefont
  {Chalkley}}, \bibinfo {author} {\bibfnamefont {S.}~\bibnamefont {Aston}},
  \bibinfo {author} {\bibfnamefont {C.~C.}\ \bibnamefont {Collins}}, \bibinfo
  {author} {\bibfnamefont {M.}~\bibnamefont {Nelson}}, \ and\ \bibinfo {author}
  {\bibfnamefont {C.~C.}\ \bibnamefont {Speake}},\ }in\ \href@noop {} {\emph
  {\bibinfo {booktitle} {Rencontres de Moriond and GPhyS Colloqium 2011}}}\
  (\bibinfo {year} {2011})\BibitemShut {NoStop}%
\bibitem [{\citenamefont {Huang}\ \emph {et~al.}(1988)\citenamefont {Huang},
  \citenamefont {Stott}, \citenamefont {Green},\ and\ \citenamefont
  {Beck}}]{Standard1}%
  \BibitemOpen
  \bibfield  {author} {\bibinfo {author} {\bibfnamefont {S.~M.}\ \bibnamefont
  {Huang}}, \bibinfo {author} {\bibfnamefont {A.~L.}\ \bibnamefont {Stott}},
  \bibinfo {author} {\bibfnamefont {R.~G.}\ \bibnamefont {Green}}, \ and\
  \bibinfo {author} {\bibfnamefont {M.~S.}\ \bibnamefont {Beck}},\ }\href
  {\doibase 10.1088/0022-3735/21/3/001} {\bibfield  {journal} {\bibinfo
  {journal} {Journal of Physics E: Scientific Instruments}\ }\textbf {\bibinfo
  {volume} {21}},\ \bibinfo {pages} {242} (\bibinfo {year} {1988})}\BibitemShut
  {NoStop}%
\bibitem [{\citenamefont {Marioli}, \citenamefont {Sardini},\ and\
  \citenamefont {Taroni}(1993)}]{Standard2}%
  \BibitemOpen
  \bibfield  {author} {\bibinfo {author} {\bibfnamefont {D.}~\bibnamefont
  {Marioli}}, \bibinfo {author} {\bibfnamefont {E.}~\bibnamefont {Sardini}}, \
  and\ \bibinfo {author} {\bibfnamefont {A.}~\bibnamefont {Taroni}},\ }\href
  {\doibase 10.1088/0957-0233/4/3/012} {\bibfield  {journal} {\bibinfo
  {journal} {Measurement Science and Technology}\ }\textbf {\bibinfo {volume}
  {4}},\ \bibinfo {pages} {337} (\bibinfo {year} {1993})}\BibitemShut {NoStop}%
\bibitem [{\citenamefont {Reverter}, \citenamefont {Li},\ and\ \citenamefont
  {Meijer}(2006)}]{Stable1}%
  \BibitemOpen
  \bibfield  {author} {\bibinfo {author} {\bibfnamefont {F.}~\bibnamefont
  {Reverter}}, \bibinfo {author} {\bibfnamefont {X.}~\bibnamefont {Li}}, \ and\
  \bibinfo {author} {\bibfnamefont {G.~C.~M.}\ \bibnamefont {Meijer}},\ }\href
  {\doibase 10.1088/0957-0233/17/11/004} {\bibfield  {journal} {\bibinfo
  {journal} {Measurement Science and Technology}\ }\textbf {\bibinfo {volume}
  {17}},\ \bibinfo {pages} {2884} (\bibinfo {year} {2006})}\BibitemShut
  {NoStop}%
\bibitem [{\citenamefont {Spinelli}\ and\ \citenamefont
  {Reverter}(2010)}]{Stable2}%
  \BibitemOpen
  \bibfield  {author} {\bibinfo {author} {\bibfnamefont {E.~M.}\ \bibnamefont
  {Spinelli}}\ and\ \bibinfo {author} {\bibfnamefont {F.}~\bibnamefont
  {Reverter}},\ }\href {\doibase 10.1109/TIM.2009.2024698} {\bibfield
  {journal} {\bibinfo  {journal} {IEEE Transactions on Instrumentation and
  Measurement}\ }\textbf {\bibinfo {volume} {59}},\ \bibinfo {pages} {458}
  (\bibinfo {year} {2010})}\BibitemShut {NoStop}%
\bibitem [{\citenamefont {Rich}(1983)}]{ShieldNoise}%
  \BibitemOpen
  \bibfield  {author} {\bibinfo {author} {\bibfnamefont {A.}~\bibnamefont
  {Rich}},\ }\href@noop {} {\bibfield  {journal} {\bibinfo  {journal} {Analog
  Devices}\ }\textbf {\bibinfo {volume} {17}},\ \bibinfo {pages} {8} (\bibinfo
  {year} {1983})}\BibitemShut {NoStop}%
\end{thebibliography}%

\end{document}